\documentstyle[10pt]{article}
\begin{document}
\begin{center}{\bf ON THE POSSIBILITY OF FASTER-THAN-LIGHT}\\
{\bf MOTION OF THE COMPTON ELECTRON}\\[2mm]
{G.A. Kotel'nikov}\\
RRC Kurchatov Institute, Kurchatov Sq. 1, Moscow 123182, Russia\\
E-mail: kga@electronics.kiae.ru\\[3mm]
\end{center}

\begin{abstract}
The kinematics of  Compton-effect  with  violated  invariance  of  the
velocity of light has been considered.  It has been shown that in this
case faster-than-light motion of the Compton electron is possible. The
motion  (if  it  exists  in  reality)  begins  with  the energy of the
incident $\gamma$-quantum above 360 keV.
\end{abstract}

\section{Introduction}\label{secA}
The Compton scattering is a  fundamental  effect  of  nuclear  physics
\cite{Akh69,  Shir72}. The successive description of its kinematics is
essential to  any  version  of  the  theory.  We  shall  consider  the
kinematics  of  this  effect  in  connection  with  the  violation  of
invariance of the speed of light in the  works  where  the  space-time
interval takes the form \cite{Kot03} - \cite{Kot04a}:
\begin{equation}\label{int}
ds^2=c^2dt^2-dx^2-dy^2-dz^2,
\end{equation}
where $t$ is the time; $x, y, z$ are the space variables, $|c|<\infty$
is the velocity of light considered as a variable.  It  is  seen  from
here  that  in  the  space with the metric (\ref{int}) the event point
coordinates are the five numbers:  the time $t$,  the space  variables
$x,  y, z,$ and the velocity of light $c$. Let us denote this space by
$V^5(t,{\bf x},c)$. In view of the absence of the space-time variables
in an explicit form in front of the differentials $dt,  dx,  dy,  dz$,
the 3-space $R^3({\bf x})\subset V^5(t,{\bf x},c)$ is homogeneous  and
isotropic,  the time $t$ is homogeneous. This is in agreement with the
basic properties of space and time in classical mechanics \cite{Lan58}
and  Special  Relativity  (SR)  \cite{Pau47}  -  \cite{Log82}.  Let us
suppose that on a particle trajectory the time has a similar  property
to the universal Newton time in classical physics:
\begin{equation}\label{time}
dt=dt_0\to t=t_0.
\end{equation}
As a  result  the velocity of light on the particle trajectory will be
depend on the particle velocity by the law
\begin{equation}\label{c}
\displaystyle
c=\pm c_0\sqrt{ 1+v^2/{c_0}^2 },
\end{equation}
\newpage\pagestyle{plain}\noindent
where $c_0={c'}_0=3\cdot10^{10}$ cm/s  is  the  proper  value  of  the
velocity  of  light.  The  particle   motion   perturbs   the   metric
(\ref{int}),  as a result of which the spectrum of $c$-values is given
by the inequality $(c_0\leq|c|<\infty)\subset(|c|<\infty)$. When $v\ne
0$, the  metric  (\ref{int}) admits a faster-than-light motion (at the
velocity $v>3\cdot10^{10}$  cm/s)  of  the  particle  with  real  mass
\cite{Kot03}  -  \cite{Kot04a}.  This  feature distinguishes the above
mentioned publications,  and the  present  work  from  the  well-known
theories  such  as  SR  \cite{Pau47}  -  \cite{Log82},  the  theory of
superluminal motions with imagine mass  \cite{Kir54}  -  \cite{Kir74},
the  theory  of  motion  with  anisotropic tensor of mass \cite{Kir54,
Kir74,  Saz72}, and the versions of electrodynamics with instantaneous
and retarded  interactions  \cite{Chu06}  -  \cite{Chu96}.  It  is the
purpose of the present work to study the kinematics of  Compton-effect
in  space-time with the metric (\ref{int}) taking into account formula
(\ref{time}) and the positive velocity of light (\ref{c}).

\begin{sloppypar}\section{Space-time transformations,            group
properties}\end{sloppypar}\label{secB}
\begin{sloppypar}\noindent
The expression  for  the  interval (\ref{int}),  which we write in the
form  $ds=F(x,c,dx)>0$,  \  $dx=(dt,dx,dy,dz)$,  possesses  the  signs
inherent      in      Finsler      space:     $F(x,c,-dx)=F(x,c,dx)>0,
F(x,c,kdx)=kF(x,c,dx)$,  $s=\int   F(x,\dot   x,c)dt=\int   (c^2-{\dot
x}^2)^{1/2}dt,  \  F(x,k\dot  x,kc)=kF(x,\dot x,c)$,  i.e.  $F$ is the
positively homogeneous function of degree $1$ with respect to $dt, dx,
dy, dz$, $v$ and $c$ \cite{Bus50, Ike78}. By replacing the variables
\end{sloppypar}
\begin{equation}\label{repl}
\begin{array}{l}
\displaystyle
x^0=\int\limits_0^t cd\tau, \ x^{1,2,3}=x,y,z, \ x^5=c
\end{array}
\end{equation}
\begin{sloppypar}\noindent
let us map the space $V^5(t,{\bf x},c)$ with the metric (\ref{int}) on
to the space $F^5(x^0,{\bf x},x^5)$ with the metric
\begin{equation}\label{m}
ds^2=(dx^0)^2-(dx^1)^2-(dx^2)^2-(dx^3)^2,
\end{equation}
where $x^0$  will  be  also  considered as re-determined "time" in the
case of the particle velocity $v\ne  const$.  The  components  of  the
metric  tensor  $g_{ab}=(+,-,-,-,0)$  ($a,b=0,1,2,3,5$)  of  the space
$F^5$ indicate that $F^5$,  as its subspaces with  the  metric  tensor
$g_{\mu\nu}=diag(+,-,-,-)$    ($\mu,   \nu=0,1,2,3$),   includes   the
Minkowski ${M^4}_1$-space on the hyper-plane $c=c_0$  with  the  local
time $x^0=c_0t$; the Minkowski ${M^4}_2$-space with the non-local time
(\ref{repl});  zero subspace ${R^1}_0(x^5)$,  which coincides with the
$x^5$-axis  \cite{Efi70}.  (In  the  ${M^4}_1$-space  a  point  on the
$x^0$-axis  corresponds  to  a  point  on   the   $t$-axis.   In   the
${M^4}_2$-space  a  point on the $x^0$-axis corresponds to an integral
$\int_0^t     c(\tau)d\tau$).     The     infinitesimal     space-time
transformations,   retaining   the   expression  (\ref{m})  under  the
condition (\ref{time}), take the form
\end{sloppypar}
\begin{equation}\label{lor}
{dx'}^{\mu}={L^{\mu}}_{\nu}dx^{\nu}, \
{x'}^5=x^5(1-{\bf\beta}\cdot{\bf u})/\sqrt{1-\beta^2},\
\mu, \nu=0,1,2,3.
\end{equation}
Here ${L^{\mu}}_{\nu}$  is  the  matrix  of  the  Lorentz group $L_6$,
$\beta={\bf V}/c=const$,  ${\bf u}={\bf v}/c$.  For the Lorentz  group
$L_1$ and   free   motions   in  $F^5$  and  $V^5$  the  corresponding
homogeneous integral transformations are
\begin{equation}\label{eq.19}
\begin{array}{l}
\displaystyle
{x'}^0=\frac{x^0-\beta x^1}{\sqrt{1-\beta^2}}, \
{x'}^1=\frac{x^1-\beta x^0}{\sqrt{1-\beta^2}}, \
{x'}^{2,3}=x^{2,3},                            \
\displaystyle  {x'}^5=x^5\frac{1-\beta u^1}{\sqrt{1-\beta^2}},
\end{array}
\end{equation}
\begin{equation}\label{eq.19a}
\begin{array}{l}
\displaystyle
x'=\frac{x-Vt}{\sqrt{1-V^2/c^2}}, \ t'=t, \ y'=y, \ z'=z, \
\displaystyle  
c'=c\frac{1-Vv_x/c^2}{\sqrt{1-V^2/c^2}},
\end{array}
\end{equation}
\begin{sloppypar}\noindent
where $u^1=v_x/c$,  $v_x=x/t$. They transform into itself the equation
of     the     surface     $(x^0)^2-(x^1)^2-(x^2)^2-(x^3)^2=0      \to
c^2t^2-x^2-y^2-z^2=0$  (the  zero  cone [12] in $F^5$,  the surface of
4-order in $V^5$).  In ${M^4}_1$-space the zero cone  changes  to  the
light    cone    ${c_0}^2t^2-x^2-y^2-z^2=0$.    The    transformations
(\ref{eq.19}) change to the Lorentz transformations.  The motions  are
described  by  SR  \cite{Pau47}  -  \cite{Log82}.  Let  us  denote the
generator    inducing    the    transformations    (\ref{eq.19})    by
$N_{01}=x_0\partial_1-x_1\partial_0+u^1x^5\partial_5$
($N_{01}=ct\partial_x+x\partial_t/c+(x/ct)(c\partial_c-t\partial_t)=
ct\partial_x+(x/t)\partial_c$  in the space $V^5$).  It belongs to Lee
algebra                of                the                 operators
$N_{\mu\nu}=x_{\mu}\partial_{\nu}-x_{\nu}\partial_{\mu}$,
$Q_0=(1/t)\partial_c,  \  P_0=(1/c)\partial_t,  \  Q_i=\partial_i,   \
Z=(c\partial_c-t\partial_t), \ i=1,2,3$:
\begin{equation}\label{lee1}
\begin{array}{l}
\displaystyle
[Q_{\mu},Q_{\nu}]=0;                                                \\
\displaystyle
[N_{\mu\nu},N_{\rho\sigma}]=-g_{\mu\rho}N_{\nu\sigma}+
g_{\mu\sigma}N_{\nu\rho}+g_{\nu\rho}N_{\mu\sigma}-
g_{\nu\sigma}N_{\mu\rho};                                            \\
\displaystyle
[Q_{\mu},N_{\nu\rho}]=g_{\mu\nu}Q_{\rho}-g_{\mu\rho}Q_{\nu};         \\
\displaystyle
[P_0,Q_{\nu}]=-g_{0\nu}Z/{x_0}^2;                                    \\
\displaystyle
[P_0,N_{\nu\rho}]=g_{0\nu}P_{\rho}-g_{0\rho}P_{\nu}-
(g_{0\rho}x_0-g_{00}x_{\rho})g_{0\nu}Z/{x_0}^2;                      \\
\displaystyle
[Z,Q_{\mu}]=[Z,P_0]=[Z,N_{\mu\nu}]=0;                                \\
\dotfill
\end{array}
\end{equation}
The algebra,  in  general case,  is infinitely dimensional.  As finite
subalgebras  it  includes  the  algebra  of   operators   $N_{\mu\nu}$
(isomorphic  to  Lee  algebra  of  Lorentz  group \cite{Bar72})),  the
algebra of operators $N_{\mu\nu},  Q_{\mu}$ (isomorphic to Lee algebra
of   Poincar\'e   group  \cite{Bar72}),  the  algebra  of  commutative
operators $[Q_{\mu},Q_{\nu}],  [Z,Q_\mu], [Z,P_0], [Z,N_{\mu\nu}]$. As
a result Lorentz and Poincar\'e groups arise in the theory not only in
the case the speed of light is invariant on the  hyper-plane  $c=c_0$,
but  also in the case the time is invariant within the transformations
(\ref{eq.19a}) in  the  $V^5$-space.  Poincar\'e  was  first  to  draw
attention  to  Lorentz  group  as  a  symmetry group of the light cone
equation  ${c_0}^2t^2-{\bf  x}^2=0$   on   the   hyper-plane   $c=c_0$
\cite{Poi05}.  The  space of $V^5$-type and the zero cone $c^2t^2-{\bf
x}^2=0$ were introduced in the papers \cite{Kot80, Kot81} in analyzing
symmetries  of  the  wave  equation  with  a non-invariant velocity of
light.
\end{sloppypar}

\begin{sloppypar}
Let us  restrict the consideration of algebra (\ref{lee1}) on a set of
functions $\phi=\phi(x^0,{\bf x})\subset f(x^0,{\bf x},x^5)$ and  take
into account $Z\phi=0$ in this case.  The algebra (\ref{lee1}) reduces
to  the  Lee  algebra  of   12-dimensional   group   $(P_{10},T_1){\rm
X}\Delta_1$,  where  $L_6\subset P_{10}$ involves hyperbolic rotations
on   the   planes   $(x^0,x^i)\subset   {M^4}_2$    (the    generators
$N_{0i}\subset  N_{\mu\nu}$),  $T_4$  involves  translations along the
$x^0,x^i$  axes  with  $t$=const  (the  generators  $Q_{\mu}$),  $T_1$
includes  translations  along  the  $x^0$  axis  with  $c$=const  (the
generator $P_0$),  $\Delta_1$ is the scale transformation of the $x^5$
axis  (generator  $Z=x^5\partial_5$).  By  using the Campbell-Hausdorf
formula \cite{Fus83},  it can be shown that consecutive operations  of
$Q_0$  and  $P_0$  are  equivalent  to the translation along the $x^0$
axis:
$
t'=e^{\theta Q_0}te^{-\theta    Q_0}=t+\theta    [Q_0,t]+\dots=t,    \
c'=e^{\theta  Q_0}ce^{-\theta Q_0}=c+\theta [Q_0,c]+\dots=c+\theta/t,\
c't'=ct+\theta; \ \ \         t''=e^{\phi               P_0}t'e^{-\phi
P_0}=t'+\phi[P_0,t']+\dots=t'+\phi/c',  \  c''=e^{\phi  P_0}c'e^{-\phi
P_0}=c'+\phi[P_0,c']+\dots=c', \ c''t''=c't'+\phi=ct+\xi 
$,
where $\xi=\theta+\phi$, $\theta$ and $\phi$ are the group parameters.
The presence  of the $P_0$ operator corresponds to motion with time if
the invariance of the speed  of  light  is  violated.  Thus,  that  is
impossible within Minkowski ${M^4}_1$-space on the hyper-plane $c=c_0$
is possible within the Minkowski  ${M^4}_2$-space  entering  into  the
Finsler space with metric (\ref{int}).
\end{sloppypar}

\section{Momentum, energy, equations of motion}\label{secC}
\begin{sloppypar}\noindent
Let us start from the connections between the partial derivatives:
\end{sloppypar}
\begin{equation}\label{partial}
\begin{array}{l}
\displaystyle      
\vspace{3mm}
\hspace{-22mm}
\frac{\partial}{\partial t}=\frac{\partial
x^0}{\partial t}\frac{\partial}{\partial x^0}+
\frac{\partial x^{i}}{\partial t}\frac{\partial}
{\partial x^{i}}+
\frac{\partial x^5}{\partial t}\frac{\partial}{\partial x^5}=
c\frac{\partial}{\partial x^0}
\Longrightarrow \frac{\partial}{\partial x^0}=
\frac{1}{c}\frac{\partial}{\partial t};  \\
\displaystyle 
\hspace{-22mm}
\frac{\partial}{\partial x}=\frac{\partial x^0}{\partial
x}\frac{\partial}{\partial x^0}+
\frac{\partial
x^{i}}{\partial x}\frac{\partial}{\partial x^{i}}+
\frac{\partial x^5}{\partial x}\frac{\partial}{\partial x^5}=
\frac{\partial}{\partial x^1}
\Longrightarrow \frac{\partial}{\partial x^1}=
\frac{\partial}{\partial x};             \\
\end{array}
\end{equation}
$$
\displaystyle
\frac{\partial}{\partial c}=\frac{\partial x^0}{\partial c}
\frac{\partial }{\partial x^0}+
\frac{\partial x^i}
{\partial c}\frac{\partial}{\partial x^i}+
\frac{\partial x^5}{\partial c}\frac{\partial}{\partial x^5}=
\frac{t}{c}\frac{\partial}{\partial t}+\frac{\partial}{\partial x^5}
\Longrightarrow
x^5\frac{\partial}{\partial x^5}=c\frac{\partial}{\partial c}-
t\frac{\partial}{\partial t}.
$$
\smallskip

\noindent
Here the expressions for $\partial/\partial y$ and  $\partial/\partial
z$ are  analogous  to $\partial/\partial x$.  It is assumed,  that the
velocity of light does not depend on space variables in the  range  of
interactions -  $\nabla  c=0$.  As  a  result  the  values of the type
$(\int_{0}^td\tau \partial       c/\partial        x)\partial/\partial
x^0$ vanish.  The  summing  is  made over twice repeating index.  Then
\cite{Kot04}:\\
{\bf -}  As  in  SR,  the  parameter  $\beta=V/c$  is  in the range of
$0\leq\beta<1$.\\
{\bf -} As in SR, $dx^0$ is the total differential.\\
{\bf -} Generally speaking,  the "time"  $x^0=\int_0^t  cd\tau$  is  a
functional of $c(\tau)$.\\
{\bf -}   The    condition    $\nabla    c(x^0)=0\leftrightarrow\nabla
c(t)=0$ is invariant on the trajectory of a particle.
\begin{sloppypar}
Keeping this  in  the  mind,  let  us  construct  the  theory  in  the
${M^4}_2$-space which is similar to  SR  in  the  ${M^4}_1$-space.  By
using the   relations  (\ref{partial}),  let  us  map  it  on  to  the
$V^5$-space with the metric (\ref{int}).  Following  \cite{Lan73},  we
start with the integral of action:
\end{sloppypar}
$$
\begin{array}{c}
\vspace{2mm}
\displaystyle
S={S}_m+{S}_{mf}+{S}_f=
-mc_0\int ds-\frac{e}{c_0}\int A_\mu dx^\mu-
\frac{1}{16\pi c_0}\int F_{\mu\nu}F^{\mu\nu} d^4x=\\
\displaystyle
\int\Big[-mc_0\sqrt{1-u^2}+\frac{e}{c_0}({\bf A}\cdot{\bf
u}-\phi)\Big]dx^0-\frac{1}{8\pi c_0}\int(E^2-H^2)d^3xdx^0=
\end{array}
$$
\begin{equation}\label{act}
\displaystyle
-mc_0\int ds-\frac{1}{c_0}\int A_\mu j^\mu d^4x-
\frac{1}{16\pi c_0}\int F_{\mu\nu}F^{\mu\nu} d^4x.
\end{equation}
\begin{sloppypar}\noindent
Here $m$  is  the  mass  of a particle;  $e$ is the electrical charge;
$S_m=-mc_0\int  ds=-mc_0\int\sqrt{1-u^2}dx^0=-mc_0\int(c_0/c)dx^0$  is
the    action   for   a   free   particle;   $S_f=-(1/16\pi   c_0)\int
F_{\mu\nu}F^{\mu\nu}d^4x$ is the action  for  a  free  electromagnetic
field;  $S_{mf}=-(e/c_0)\int  A_\mu  dx^\mu=-(1/c_0)\int  A_\mu  j^\mu
d^4x$ is the action corresponding to the  interaction  of  the  charge
with electromagnetic field; $A^\mu=(\phi,{\bf A})$ is the 4-potential;
$A_\mu=g_{\mu\nu}A^\nu=(\phi,-{\bf A})$; $j^\mu=(\rho,\rho{\bf u})$ is
the 4-vector of the density of a current; $\rho$ is the density of the
charge;  ${\bf u}={\bf v}/c$ is  the  dimensionless  3-velocity  of  a
particle;  ${\bf  E}=-\partial{\bf  A}/\partial x^0-\nabla\phi$ is the
electric field;  ${\bf H}=\nabla{\em X}{\bf A}$ is the magnetic field;
$F_{\mu\nu}=\partial   A_\nu/\partial   x^\mu-\partial  A_\mu/\partial
x^\nu$    is    the    tensor    of    an    electromagnetic    field;
$F_{\mu\nu}F^{\mu\nu}=2(H^2-E^2)$;   $d^4x=dx^0dx^1dx^2dx^3$   is  the
element of the invariant 4-volume.  The speed of light $c_0$, the mass
of  a particle $m$,  the electrical charge $e$ are invariant constants
of the theory.
\end{sloppypar}
In spite of the fact that the action (\ref{act})  is  similar  to  the
action of  SR,  it  differs  from  the  SR  action  \cite{Lan73}.  The
electrical    field    has     been     chosen     in     the     form
${\bf E}=-\partial{\bf  A}/\partial x^0-\nabla\phi=-(1/c)\partial {\bf
A}/\partial t-\nabla\phi$  instead  of  ${\bf E}=-(1/c_0)\partial {\bf
A}/\partial t-\nabla\phi$  \cite{Lan73}  \cite{Lan73}.   The   current
density has    been   chosen   in   the   form   $j^\mu=(\rho,\rho{\bf
u})=(\rho,\rho{\bf v}/c)$     instead     of     $j^\mu=(\rho,\rho{\bf
v})$ \cite{Lan73}.  The  current  density is similar to the expression
from Pauli monograph \cite{Pau47} with the only difference that  ${\bf
j}$ in  (\ref{act})  is  equal to $\rho{\bf v}/c$ instead of $\rho{\bf
v}/c_0$ \cite{Pau47}.  Analogously,  the propagation velocity  of  the
4-potential in   (\ref{act})   is   equal  to  $c$  instead  of  $c_0$
\cite{Lan73}. The action (\ref{act}) goes into the SR  action,  if  we
replace $c$   by   $c_0$  within  the  corresponding  expressions.  In
accordance with the construction,  the action (\ref{act})  is  Lorentz
invariant and  does not depend on the $x^5$ variable.  As a result the
action (\ref{act})  is   invariant   with   respect   to   the   group
$(P_{10},T_1){\rm X}\Delta_1$, induced by the reduction of the algebra
(\ref{lee1}) on the set of  functions  $\phi=\phi(x^0,{\bf  x})$.

Lagrangian $L$, the generalized momentum ${\bf P}$ and the generalized
energy ${\cal H}$ take the form:
\begin{equation}\label{mom}
\displaystyle
L=-mc_0\sqrt{1-u^2}+\frac{e}{c_0}({\bf A}\cdot{\bf u}-\phi);
\end{equation}
\begin{equation}\label{imp}
\displaystyle
{\bf P}=\frac{\partial  L}{\partial{\bf u}}=\frac{mc_0{\bf u}}{\sqrt
{1-u^2}}+\frac{e}{c_0}{\bf A}={\bf p}+\frac{e}{c_0}{\bf A}=m{\bf v}+
\frac{e}{c_0}{\bf A};
\end{equation}
\begin{equation}\label{ener}
\displaystyle
{\cal H}={\bf P}\cdot{\bf u}-L=\frac{mc_0c+e\phi}{c_0}=
\frac{{\cal E}+e\phi}{c_0}.
\end{equation}
\begin{equation}\label{pe}
\displaystyle
{\bf p}=m{\bf v}, \ {\cal
E}=mcc_0=m{c_0}^2\sqrt{1+\frac{v^2}{{c_0}^2}}.
\end{equation}
Here ${\bf  p}$,  ${\cal E}$ are the momentum and energy of a particle
with mass $m$. They may be combined into 4- momentum $p^\mu$
\begin{equation}\label{p}
p^\mu=mc_0u^\mu=\Big(\frac{mc_0c}{c_0},mcu^i\Big)=
\Big(\frac{{\cal E}}{c_0},m{\bf v}\Big),
\end{equation}
the components of which are related as follows:
\begin{equation}\label{dis}
\displaystyle
p_\mu p^\mu=\frac{{\cal   E}^2}{{c_0}^2}-{\bf    p}^2={m}^2{c_0}^2;   \ \
{\bf   p}=\frac{{\cal   E}}{c_0c}{\bf   v}; \
{\bf p}=\frac{{\cal E}}{c_0c}{\bf c}, \ if \ m=0, \ {\bf v}={\bf c}.
\end{equation}
It is seen from here that the momentum of a  particle  with  the  mass
$m=0$  is  independent  of the absolute value of the particle velocity
$v=c$.  It is determined only by  the  energy  of  a  particle:  ${\bf
p}={\bf n}{\cal E}/c_0,  \ {\bf n}={\bf c}/c$. (In SR this property is
masked by $c_0$ being constant).

Next, let us start from the  mechanical  \cite{Lan73}  and  the  field
equations \cite{Bog73} of Lagrange
\begin{equation}\label{Lagr}
\displaystyle
\frac{d}{dx^0}\frac{\partial L}{\partial{\bf u}}-
\frac{\partial L}{\partial{\bf x}}=0; \
\frac{\partial}{\partial x^\nu}\frac{\partial {\cal L}}
{\partial(\partial
A_\mu/\partial x^\nu)}-\frac{\partial {\cal L}}{\partial A_\mu}=0,
\end{equation}
\begin{sloppypar}\noindent
where $L$  is  the  Lagrangian  (\ref{mom}),  ${\cal  L}=-(1/c_0)A_\mu
j^\mu-(1/16\pi c_0) F_{\mu\nu}F^{\mu\nu}$ is the density  of  Lagrange
function  for  electromagnetic field and interaction between the field
and the charge.  Taking into consideration the  equality  $\nabla({\bf
a}\cdot{\bf  b})= ({\bf a}\cdot\nabla){\bf b}+({\bf b}\cdot\nabla){\bf
a}+{\bf a}{\rm x} (\nabla{\rm  x}{\bf  b})+{\bf  b}{\rm  x}(\nabla{\rm
x}{\bf a})$,   the   permutable   relationships   for  the  tensor  of
electromagnetic field,                 the                  expression
$\partial(F_{\mu\nu}F^{\mu\nu})/\partial(\partial       A_\mu/\partial
x^\nu)= -4F^{\mu\nu}$ \cite{Lan73},  we find the equations of  motions
of electromagnetic field and of a particle in the field
\begin{equation}\label{equ}
\begin{array}{c}
\vspace{2mm}
\displaystyle
\frac{d{\bf p}}{dx^0}=\frac{e}{c_0}{\bf E}+
\frac{e}{c_0}{\bf  u}{\rm  x}{\bf  H};      \
\frac{d{\cal E}}{dx^0}=e{\bf E}\cdot{\bf u};\\
\vspace{2mm}
\displaystyle
\frac{\partial F_{\mu\nu}}{\partial x^\rho}+
\frac{\partial F_{\nu\rho}}{\partial x^\mu}+
\frac{\partial F_{\rho\mu}}{\partial x^\nu}=0; \
\displaystyle
\frac{\partial F^{\mu\nu}}{\partial x^\nu}+4\pi  j^\mu=0.
\end{array}
\end{equation}
(Here ${\bf       p}=mc_0{\bf       u}/\sqrt{1-u^2},      \      {\cal
E}=m{c_0}^2/\sqrt{1-u^2}$).  In  the   variables   $(x^0,x^1,x^2,x^3)$
equations (\ref{equ}) coincide exactly with the equations \cite{Lan73}
and are the same  for  both  the  Minkowski  spaces  -  ${M^4}_1$  and
${M^4}_2$. The difference arises if the equations are written with the
variables $(t,x,y,z)$.  In the case of ${M^4}_1$-Minkowski  space  the
equations coincide with SR equations \cite{Lan73}, if we put $c=c_0, \
dx^0=c_0dt$ into  them.  (In  accordance   with   going   the   action
(\ref{act}) into   the   SR  action  \cite{Lan73}).  In  the  case  of
${M^4}_2$-Minkowski space  it  is  necessary  to  take  into   account
$dx^0=cdt, \  \sqrt{1-u^2}=c_0/c$,  and the relations (\ref{partial}).
Then the  equations  of  motions  take  the   forms   \cite{Kot03}   -
\cite{Kot04a}:
\end{sloppypar}
\begin{equation}\label{mov}
\displaystyle
\frac{d{\bf p}}{dt}=m\frac{d{\bf v}}{dt}=\frac{c}{c_0}e{\bf E}+
\frac{e}{c_0}{\bf v}{\rm x}{\bf H}; \
\frac{d{\cal E}}{dt}=e{\bf E}\cdot{\bf v}\to m\frac{dc}{dt}=
\frac{e}{c_0}{\bf v}\cdot{\bf E}.
\end{equation}
\begin{equation}\label{max}
\begin{array}{ll}
\vspace{2mm}
\displaystyle
\nabla{\rm X}{\bf E}+\frac{1}{c}\frac{\partial{\bf H}}{\partial t}=0; &
\displaystyle
\nabla\cdot{\bf E}=4\pi\rho;                                         \\
\displaystyle
\nabla{\rm X}{\bf H}-\frac{1}{c}\frac{\partial{\bf E}}{\partial t}=
4\pi\rho\frac{{\bf v}}{c};                                            &
\nabla\cdot{\bf H}=0,
\end{array}
\end{equation}
\begin{sloppypar} \noindent
where $c(t)=c_0(1+v^2/{c_0}^2)^{1/2}=c(0)[1+(e/mc_0c(0))  \int_0^t{\bf
v}\cdot{\bf   E}d\tau]$,   $\nabla   c=0$.   Equations  (\ref{mov})  -
(\ref{max}),  if  considered  as  the  whole,  form  a  set   of   the
self-consistent    nonlinear    equations.   (In   the   approximation
$v^2/{c_0}^2\ll 1$ by $c\sim c_0$,  \  they  describe  the  motion  of
non-relativistic  particle  in electromagnetic field and coincide with
\cite{Lev69}).  They admit faster-than-light motion of a particle with
the real mass $m$, rest energy ${\cal E}_0=m{c_0}^2$ and the velocity
\end{sloppypar}
\begin{equation}\label{v}
v=\sqrt{{\cal E}^2-{m}^2{c_0}^4}/mc_0>c_0,
\end{equation}
if the energy of a particle satisfies the inequality ${\cal  E}>\sqrt2
{\cal E}_0$.  For  example,  for  the  proton the rest energy is equal
$938$~MeV. The  1~GeV   proton   velocity   is   about   $0.37   c_0$.
Faster-than-light motion  of  the  proton  begins with the energy $\sim
1.33$~ GeV.  The faster-than-light electron motion  (${\cal  E}_0=511$
keV) begins with the energy $\sim 723$ keV. The calculated velocity of
1 GeV electron is $\sim~2000~c_0$.  Thus, if ${M^4}_2$-Minkowski space
were  realized in the nature,  the neutron physics of nuclear reactors
could be formulated in the approximation $v\ll c_0$,  as  in  SR.  The
particle  physics  on  modern  accelerators  would  be  the physics of
faster-than-light motions.  The results obtained are given in Table  1
in  comparison with the analogous results from classical mechanics and
SR. In   this   Table   the   designations    are    used:    $d{\bf
x}^2=dx^2+dy^2+dz^2$,   $T$   is   the  kinetic  energy,  $\beta=V/c$.

\hspace{100mm} Table 1                                             \\
$$
\begin{array}{|c|c|c|}\hline\hline
{}&{}&{}\\
{}The \ classical \ mechanics{}& Special \ Relativity               &
Present \ work                                                      \\
\cite{Lan58}  &{\cite{Pau47} - \cite{Log82}}                        &
\cite{Kot03} - \cite{Kot04a}                                        \\
{}&{}&{}\\
\hline
{}&{}&{}                                                            \\
ds^2=d{\bf x}^2 & ds^2={c_0}^2dt^2-d{\bf x}^2                       &
ds^2=c^2dt^2-d{\bf x}^2                                             \\
{}&{}&{}\\
\hline
x'=x-Vt, &
\displaystyle 
x'=\frac{x-Vt}{\sqrt{1-\beta^2}},                                   &
\displaystyle 
x'=\frac{x-Vt}{\sqrt{1-\beta^2}},                                   \\
y'=y, \ z'=z, & y'=y, \ z'=z, & y'=y, \ z'=z,                       \\
t'=t, &
\displaystyle 
t'=\frac{t-Vx/{c_0}^2}{\sqrt{1-\beta^2}},                           &
t'=t,                                                               \\
c'=c\sqrt{1-2\beta n_x+\beta^2}                                     &
{c_0}'=c_0                                                          &
\displaystyle
c'=c\frac{1-Vv_x/c^2}{\sqrt{1-\beta^2}}                             \\
\hline
{\bf p}=m{\bf v}                                                    &
\displaystyle
{\bf p}=\frac{m{\bf v}}{\sqrt{1-v^2/{c_0}^2}}                       &
{\bf p}=m{\bf v}                                                    \\
\displaystyle
T=\frac{mv^2}{2}                                                     &
\displaystyle
{\cal E}=\frac{m{c_0}^2}{\sqrt{1-v^2/{c_0}^2}}                       &
\displaystyle
{\cal E}=m{c_0}^2{\sqrt{1+v^2/{c_0}^2}}                             \\
\displaystyle
T=\frac{p^2}{2m}                                                    &
\displaystyle
{\cal E}^2-{c_0}^2p^2=m^2{c_0}^4                                    &
\displaystyle
{\cal E}^2-{c_0}^2p^2=m^2{c_0}^4                                    \\
\displaystyle
m\frac{d{\bf v}}{dt}=e{\bf E}+\frac{e}{c_0}{\bf v}{\rm x}{\bf H}
\cite{Lev69}                                                        &
\displaystyle
\frac{d{\bf p}}{dt}=e{\bf E}+\frac{e}{{c_0}}{\bf v}{\rm x}{\bf H}   &
\displaystyle
m\frac{d{\bf v}}{dt}=
\frac{c}{{c_0}}e{\bf E}+\frac{e}{{c_0}}{\bf v}{\rm x}{\bf H}        \\
\displaystyle
\frac{dT}{dt}=e{\bf v}\cdot{\bf E}                                  &
\displaystyle
\frac{d{\cal E}}{dt}=e{\bf v}\cdot{\bf E}                           &
\displaystyle
\frac{d{\cal E}}{dt}=e{\bf v}\cdot{\bf E}                           \\
{}&{}&{}\\
\hline\hline
\end{array}
$$

It is shown in \cite{Kot03} - \cite{Kot04a},  how a lot of experiments
(interpreted only by SR until the present time) may be explained  with
the   help  of  the  proposed  theory.  For  example,  these  are  the
experiments  of  Michelson  and  Fizeau,  aberration  of  light,   the
appearance   of   atmospheric   $\mu$-mesons  on  the  Earth  surface,
Doppler-effect,  a number of the known experiments for  the  proof  of
independence of the speed of light from the emitter velocity, decay of
unstable particles,  generation of new particles in nuclear reactions,
Compton-effect,   photo-effect.   We   consider   the   kinematics  of
Compton-effect in more detail.

\section{Motion integrals: momentum and energy}
\begin{sloppypar}\noindent
Let us note that $dx^0$,  according to the construction,  is the total
differential. As a result  $x^0$  possess  property of  the  time  for
${M^4}_2$-Minkowski space. Therefore, by virtue of Lagrange mechanical
equations, the momentum and energy (\ref{pe}) for an  isolated  system
are  the  integrals of motion because of the homogeneity of space-time
\cite{Lan58, Lan73}. The formula for the kinetic energy takes the form
\end{sloppypar}
\begin{equation}\label{cin}
\displaystyle
T={\cal E}-m{c_0}^2=m{c_0}^2(\frac{c}{c_0}-1)=
m{c_0}^2\Big(\sqrt{1+\frac{v^2}{{c_0}^2}}-1\Big)\approx
\frac{1}{2}mv^2.
\end{equation}
With $v^2\ll   {c_0}^2$  expression  (\ref{cin})  coincides  with  the
expression for kinetic energy  in  classical  mechanics  (as  in  SR).
Variations of  ${\cal  E}$  and  ${\bf  p}$  with  time  determine the
dynamics of a particle for  ${M^4}_2$-Minkowski  space.  With  $v^2\ll
{c_0}^2$ the  new  dynamics goes into the Newton dynamics.

Let us  use the expressions (\ref{pe}) and (\ref{cin}) to describe the
motion of a lot of number of particles.  Following  \cite{Shir72},  we
shall  consider  the  reaction  in  which  the  particles  with masses
${m'}_1,  {m'}_2,  \dots, {m'}_n$ are produced in colliding the moving
particle  $m_1$ with the immobile particle $m_2$ (the target).  Let us
write the conservation laws in the form
\begin{equation}\label{sys}
\begin{array}{c}
{\bf p}_1={\bf p'}_1+{\bf p'}_2+ \dots + {\bf p'}_n,\\
{\cal E}_1+m_2{c_0}^2={\cal E'}_1+{\cal E'}_2+\dots+{\cal E'}_n,
\end{array}
\end{equation}
where the momentum and energy of each of the particle are given by the
formulas   (\ref{pe})   (${\bf   p'}_i={m'}_i{\bf   v'}_i,   \   {\cal
E'}_i={m'}_i{c'}_ic_0$). By   using   the   relationship  between  the
momentum and    energy    ${{\cal     E}_1}^2={c_0}^2{{\bf     p}_1}^2
+{m_1}^2{c_0}^4$ and  the  property  of  invariance  of the expression
$(\sum_i{\cal E}_i)^2-{c_0}^2(\sum_i{\bf p}_i)^2=inv$,  we  may  write
the expression    of   the   threshold   energy   ${\cal   E}_{1,thr}$
of reaction (\ref{sys}) in the form
\begin{equation}\label{tr}({\cal
E}_{1,thr}+m_2{c_0}^2)^2-{c_0}^2{{\bf p}_1}^2=(\sum_i
{m'}_i)^2{c_0}^4.
\end{equation}
From here we find the threshold kinetic energy
\begin{equation}\label{cine}
\displaystyle
T_{1,thr}=\frac{(\sum_i{m'}_i+m_1+m_2)(\sum_i{m'}_i-m_1-m_2)}{2m_2}
{c_0}^2.
\end{equation}
It coincides  with  the  similar  formula  from SR \cite{Shir72}.  The
difference arises in calculating the velocity of  a  hitting  particle
and  the  threshold  velocity  of  the reaction products.  Taking into
account ${\cal E}_{1,thr}=T_{1,thr}+m_1{c_0}^2$, we find the threshold
velocity of the particle $m_1$:
\begin{equation}\label{pp}
\displaystyle
v_1=c_0\sqrt{\Big[1+\frac{(\sum_i{m'}_i+m_1+m_2)(\sum_i{m'}_i-m_1-m_2)}
{2m_1m_2}\Big]^2-1}.
\end{equation}
It follows  from the momentum-energy conservation law (\ref{sys}) that
the velocity of the conglomerate of particles $\sum_i{m'}_i$ moving at
the same (threshold) velocity $V'$, will be equal
\begin{equation}\label{V}
\displaystyle
V'=\frac{p_1}{\sum_i{m'}_i}=\frac{m_1}{\sum_i{m'}_i}v_1
\end{equation}
In the case of proton-proton collision $p^++p^+=p^++p^++p^++p^-$, when
$m_1=m_2={m'}_i=m_p$, we obtain that the threshold energy for creating
the antiproton  is  equal   ${\cal   E}_{1,thr}=7m_p{c_0}^2\sim   6.6$
GeV in  accordance  with \cite{Shir72},  and $v_1=\sqrt{48}c_0\sim 6.9
c_0$, $V'=\sqrt3c_0\sim  1.7c_0$  in  accordance  with   \cite{Kot04}.
In SR  these  values  are equal ${\cal E}_{1,thr}=7m_p{c_0}^2\sim 6.6$
GeV \cite{Shir72},  $v_1\to w_1= \sqrt{48/49}c_0\sim 0.99c_0$,  $V'\to
W'=\sqrt{3/4}c_0\sim 0.87c_0$ respectively.

\begin{sloppypar}\section{Consequences of momentum-energy conservation
law for Compton-effect}
\end{sloppypar}\noindent
Let us  consider  the  kinematics  of $\gamma$-quantum scattering on a
free electron with the rest energy ${\cal E}_0=m{c_0}^2$, where $m$ is
the  mass  of electron.  By using the momentum-energy conservation law
and without concretizing the expressions for the momentum  ${\bf  p}'$
and energy ${\cal E}'$ of the scattered electron, we find
\begin{equation}\label{eq.1}
\begin{array}{l}
\displaystyle
\hbar\omega+{\cal E}_0=\hbar\omega'+{\cal E}';                        \\
\displaystyle
\frac{\hbar\omega}{c_0}=\frac{\hbar\omega'}{c_0}cos\theta+p'cos\alpha;\\
\displaystyle
0=\frac{\hbar\omega'}{c_0}sin\theta-p'sin\alpha.
\end{array}
\end{equation}
Here $\hbar$ is the Planck constant,  $\omega$ and $\omega'$  are  the
frequencies of    the    incident   and   scattered   $\gamma$-quanta,
$\hbar\omega$ and $\hbar\omega'$ are the energies of these quanta. The
momentum  of  the  incident  $\gamma$-quantum  is  directed  along the
$\alpha$   $x$-axis,   $\theta$   is   the   scattering    angle    of
$\gamma'$-quantum,  $\alpha$ is the scattering angle of electron $e'$.
The angle $\theta$ is counted counterclockwise;  the angle $\alpha$ is
counted clockwise. Let us rewrite the momentum conservation law in the
form
\begin{equation}\label{eq.2}
\begin{array}{l}
\displaystyle
{p'}^2cos^2\alpha=
\Big(\frac{\hbar\omega}{c_0}-\frac{\hbar\omega'}{c_0}cos\theta\Big)^2, \
\displaystyle
{p'}^2sin^2\alpha=
\Big(\frac{\hbar\omega'}{c_0}\Big)^2sin^2\theta,
\end{array}
\end{equation}
\begin{sloppypar}\noindent
and square  this.  By  summing  the  result  obtained and by using the
conservation energy law and the dispersion expression (\ref{dis}),  we
find the  known  formula  for  the scattered $\gamma'$-quantum angular
distribution and its frequency \cite{Akh69, Shir72}
\end{sloppypar}
\begin{equation}\label{eq.3}
\displaystyle
\omega'=\frac{\omega}{1+\frac{\hbar\omega}{{\cal E}_0}(1-cos\theta)}.
\end{equation}
With the  help of (\ref{eq.3}) we find the scattered $\gamma'$-quantum
momentum:
\begin{equation}\label{eq.3a}
\displaystyle
{\bf p'}_{\gamma}=\frac{\hbar\omega'}{c_0}(cos\theta,sin\theta)=
\frac{\hbar\omega}{c_0[1+\frac{\hbar\omega}{{\cal E}_0}(1-cos\theta)]}
(cos\theta,sin\theta).
\end{equation}

The scattered electron momentum may  be  found  by  means  of  putting
(\ref{eq.3}) into (\ref{eq.2}):
\begin{equation}\label{eq.10}
\begin{array}{l}
\vspace{1mm}
\displaystyle
p'cos\alpha=\frac{\hbar\omega}{{\cal E}_0}mc_0
\frac{({\cal E}_0+\hbar\omega)(1-cos\theta)}{{\cal
E}_0+\hbar\omega(1-cos\theta)}; \\
\displaystyle
p'sin\alpha=\frac{\hbar\omega}{{\cal E}_0}mc_0
\frac{{\cal E}_0sin\theta}{{\cal E}_0+\hbar\omega(1-cos\theta)}.
\end{array}
\end{equation}
\begin{equation}\label{eq.11}
\begin{array}{l}
\vspace{1mm}
\displaystyle
p'(\theta)=\frac{\hbar\omega}{{\cal E}_0}mc_0
\frac{\sqrt{{{\cal E}_0}^2sin^2\theta+({\cal
E}_0+\hbar\omega)^2(1-cos\theta)^2}}{{\cal E}_0+\hbar\omega(1 -
cos\theta)}.
\end{array}
\end{equation}
The relationship  between  the  scattered  electron  angle   and   the
scattered $\gamma'$-quantum     angle     may    be    derived    from
(\ref{eq.10}) and takes the form
\begin{equation}\label{eq.12a}
\displaystyle
tg\alpha=\frac{ {\cal E}_0 sin\theta }
{({\cal E}_0+\hbar\omega)(1-cos\theta)}.
\end{equation}
The equality $\alpha=0$ induces  the  solutions  $\theta=\pm  k\pi,  \
k=0,1,2,\dots$, which  corresponds  to  propagation  of  the scattered
$\gamma'$-quantum along and  opposite  the  direction  of  moving  the
Compton electron.  Suppose  $\theta=0$  and $\theta=\pi$,  we find the
expressions for the forward  scattered  electron  momentum  $p'$  with
$\alpha=0$:
\begin{equation}\label{eq.6}
\begin{array}{llll}
\alpha=0; &\theta=0;  &\omega'=\omega; & {p'}_{\theta=0}=0; \\
\alpha=0; &\theta=\pi;&
\displaystyle
\omega'=\frac{\omega}{1+\frac{ 2\hbar\omega}{{\cal E}_0}};&
\displaystyle
{p'}_{\theta=\pi}=
\frac{\hbar\omega}{{\cal E}_0}mc_0\Big[1+\frac{{\cal E}_0}
{{\cal E}_0+2\hbar\omega}\Big].
\end{array}
\end{equation}

The Compton-electron  energy  may  be  found  by  putting  the formula
(\ref{eq.11}) into the dispersion relationship (\ref{dis}):
\begin{equation}\label{eq.E}
\begin{array}{c}
\vspace{1mm}
\displaystyle
{\cal E}'=\sqrt{{{\cal E}_0}^2+\hbar^2\omega^2
\frac{{{\cal E}_0}^2sin^2\theta+
({\cal  E}_0+\hbar\omega)^2(1-cos\theta)^2  }
{[{\cal E}_0+\hbar\omega(1-cos\theta)]^2  }    }=\\
\vspace{1mm}
\displaystyle
{\cal E}_0+\frac{{\hbar}^2\omega^2(1-cos\theta)}{{\cal E}_0
+\hbar\omega(1-cos\theta)}.
\end{array}
\end{equation}
\begin{sloppypar}\noindent
The second,  simple form of this formula  was  derived  by  using  the
energy conservation law (\ref{eq.1}) taking into account the frequency
$\omega'$ from  (\ref{eq.3}).  It  is  essential  that all the results
obtained are independent of concrete expressions for  the  energy  and
momentum (${\cal E}=m{c_0}^2/ \sqrt{1-v^2/{c_0}^2}$,  \ ${\bf p}=m{\bf
v}/\sqrt{1-v^2/{c_0}^2}$ for             ${M^4}_1$;             ${\cal
E}=m{c_0}^2\sqrt{1+v^2/{c_0}^2}$, \ ${\bf p}=m{\bf v}$ for ${M^4}_2$).
Therefore, in view of the laws of conservation  (\ref{eq.1})  and  the
dispersion relationship  (\ref{dis}),  these  are  common  to both the
Minkowski spaces.  The distinctions arise  when  the  transformational
properties of  the time "t" for the ${M^4}_1$ and ${M^4}_2$-spaces are
taken into account in calculating  the  velocities  of  the  scattered
$\gamma'$-quantum and    Compton    electron.    By    using   formula
(\ref{eq.11}), we  find  that  in  the  ${M^4}_2$-space  the   Compton
electron velocity is
\end{sloppypar}
\begin{equation}\label{eq.13}
\begin{array}{l}
\vspace{1mm}
\displaystyle
{v'}(\theta)=\frac{\hbar\omega}{{\cal E}_0}c_0
\frac{\sqrt{{{\cal E}_0}^2sin^2\theta+({\cal
E}_0+\hbar\omega)^2(1-cos\theta)^2}}{{\cal E}_0+\hbar\omega(1 -
cos\theta)}.
\end{array}
\end{equation}
It is equal to zero with $\theta=0$.  When $\theta=\pi$,  the electron
velocity will  exceed  the  speed  of  light  $c_0$  if  the following
inequality holds:
\begin{equation}\label{eq.7}
\displaystyle
{v'}(\alpha=0,\theta=\pi)=
c_0\frac{\hbar\omega}{{\cal E}_0}\frac{2({\cal
E}_0+\hbar\omega)}{{\cal E}_0+2\hbar\omega}>c_0.
\end{equation}
According to    (\ref{eq.7}),    faster-than-light   motion   of   the
forward-scattered electron begins from  the  energy  of  the  incident
$\gamma$-quantum:
\begin{equation}\label{eq.8}
\displaystyle
\hbar\omega>\frac{{\cal E}_0}{\sqrt2}\sim 360 \ {\rm keV}.
\end{equation}

\begin{sloppypar}
Thus, it  follows  from  the  kinematics  of  Compton-effect  that  in
scattering the $\gamma$-quantum in the ${M^4}_2$-Minkowski space,  the
appearance of electron  faster-than-light  motion  is  possible.  This
motion begins  from  the  $\gamma$-quantum energy exceeding $360$ keV.
For going to SR, the following relations may be used:
\begin{equation}\label{sr-pw}
\displaystyle
v'=\frac{w'}{\sqrt{1-{w'}^2/{c_0}^2}}, \
w'=\frac{v'}{\sqrt{1+{v'}^2/{c_0}^2}},
\end{equation}
where $v'$ is the velocity of Compton electron in the ${M^4}_2$-space,
$w'$ is the velocity of Compton electron in the ${M^4}_1$-space:
\begin{equation}\label{vsr}
\begin{array}{c}
\displaystyle
{w'}(\theta)=
\frac{\hbar\omega}{{\cal E}_0}c_0\sqrt{\frac{{{\cal E}_0}^2}
{\hbar^2\omega^2+\frac{{{\cal E}_0}^2[{\cal E}_0+\hbar\omega(1-cos\theta)]^2}
{[{{\cal E}_0}^2sin^2\theta+({\cal E}_0+\hbar\omega)^2(1-cos\theta)^2]}}}<c_0.
\end{array}
\end{equation}
The relations   (\ref{sr-pw})   correspond  to  the  equality  of  the
scattered electron energy in both the Minkowski spaces.
\end{sloppypar}

To calculate  the  scattered   $\gamma'$-quantum   velocity   in   the
${M^4}_2$-space,  the  use  of the energy-momentum conservation law is
scarce. Certain assumptions of the nature of scattering are necessary.
\bigskip

\begin{sloppypar}
\section{The possible mechanisms of Compton scattering}
\end{sloppypar}
\noindent
\subsection{Local scattering}
\begin{sloppypar}\noindent
Suppose, an  incident  $\gamma$-quantum  is  scattered  by an immobile
electron in the point of  its  localization  in  accordance  with  the
Feynman diagram        corresponding        to       the       process
\end{sloppypar}
\bigskip

\begin{picture}(5,5)
$\gamma$
\put(0,0){\line(1,-2){10}}
\put(10,-20){\line(1,2){10}}
\put(0,-40){\thicklines\line(1,2){10}}
\put(10,-20){\thicklines\line(1,-2){10}}
\hspace{7mm}$\gamma'$
\hspace{20mm}$\gamma+e^-\to\gamma' +{e^-}'.$
\end{picture}
\vspace{20mm}

\noindent (The thin lines correspond to $\gamma$-quanta, the bold line
correspond to electrons). As a result of the interaction the scattered
$\gamma'$-quantum velocity  will  be equal $c'=c_0=3\cdot10^{10}$ cm/s
and independent of the scattering angle $\theta$  in  accordance  with
$c'=c_0\sqrt{1+v^2/{c_0}^2}$, if    $v=0$.   The   forward   scattered
$\gamma'$-quantum velocity ($\theta=0$), as well as the back scattered
$\gamma'$-quantum velocity ($\theta=\pi$) will be equal the same value
of $3\cdot10^{10}$   sm/s.   The   electron   gains    the    velocity
(\ref{eq.13}), becoming the electron ${e^-}'$.

\subsection{Non-local scattering. Version A}
According to  quantum  electrodynamics concepts \cite{Akh69,  Shir72},
suppose  the  scattering  is  described   by   the   Feynman   diagram
corresponding to the process 
\bigskip

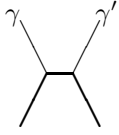
\begin{picture}(5,5)
$\gamma$
\put(0,0){\line(1,-2){10}}
\put(10,-20){\thicklines\line(2,0){10}}
\put(20,-20){\line(1,2){10}}
\put(0,-40){\thicklines\line(1,2){10}}
\put(20,-20){\thicklines\line(1,-2){10}}
\hspace{10mm}$\gamma'$
\hspace{20mm}$\gamma+e^-\to(e^-)_v\to\gamma' +{e^-}'.$
\end{picture}
\vspace{20mm}

\noindent
The incident  $\gamma$-quantum is absorbed by the immobile electron in
some point of space-time, after which an intermediate state is formed,
the virtual  electron  ${e^-}_v$.  Next the virtual electron emits the
$\gamma'$-quantum in another point of space-time and becomes the  free
scattered electron  ${e^-}'$.  By  determining the electron mass $m_v$
and the virtual  electron  velocity  $v_v$  from  the  energy-momentum
conservation law
\footnote{The full set of equations with participation of the
virtual electron may be written as follows:
$$\label{system}
\begin{array}{l}
\displaystyle
\hbar\omega+{\cal E}_0=m_v{c_0}^2(1+{v_v}^2/{c_0}^2)^{1/2}, \
\displaystyle
\hbar\omega/{c_0}=m_vv_v;                                   \\
\displaystyle
m_v{c_0}^2(1+{v_v}^2/{c_0}^2)^{1/2}=
\hbar\omega'+{\cal E}_0(1+{v'}^2/{c_0}^2)^{1/2};            \\
\displaystyle
m_vv_v=({\hbar\omega'}/{c_0})cos\theta+p'cos\alpha;         \
\displaystyle
0=({\hbar\omega'}/{c_0})sin\theta-p'sin\alpha.
\end{array}
$$
By eliminating  $m_v{c_0}^2(1+{v_v}^2/{c_0}^2)^{1/2}$  and   $m_vv_v$,
this set may be reduced to set (\ref{eq.1}).}
\begin{equation}\label{eq.12}
\begin{array}{l}
\displaystyle
\hbar\omega+{\cal E}_0=m_v{c_0}^2(1+{v_v}^2/{c_0}^2)^{1/2}, \
\displaystyle
\hbar\omega/{c_0}=m_vv_v,
\end{array}
\end{equation}
we find the expression for the  virtual  electron  mass  and  the  its
velocity
\begin{equation}\label{eq.13a}
\begin{array}{c}
\displaystyle
m_v=\frac{\sqrt{{{\cal E}_0}^2+2\hbar\omega{\cal E}_0}}{{c_0}^2};   \\
\displaystyle
v_v=c_0\frac{\hbar\omega}{\sqrt{{{\cal       E}_0}^2+2\hbar\omega{\cal
E}_0}}\sim c_0\sqrt{\frac{\hbar\omega}{2{\cal E}_0}}, {\rm if}    \
\hbar\omega\gg{\cal E}_0.
\end{array}
\end{equation}
The scattered $\gamma'$-quantum velocity will be equal
\begin{equation}\label{eq.A}
\displaystyle
c'=c_0\sqrt{1+\frac{{v_v}^2}{{c_0}^2}}=
c_0\sqrt{ 1+\frac{\hbar^2\omega^2}{{{\cal E}_0}^2+2\hbar\omega{\cal
E}_0}}.
\end{equation}
It does   not   depend   on   the   scattering   angle   $\theta$   of
$\gamma'$-quantum and with $\hbar\omega\gg{\cal E}_0$ is equal $c'\sim
c_0\sqrt{\hbar\omega/2{\cal E}_0}>c_0$.

\subsection{Non-local scattering. Version B}
\begin{sloppypar}\noindent
Let us  note  that the set of equations with a virtual electron admits
another mechanism of Compton scattering.  Suppose the virtual electron
${e^-}_v$ transmutes  spontaneously  into  the  free electron ${e^-}'$
that emits the $\gamma'$-quantum (the scattered gamma-quantum). In the
${M^4}_1$-space both  the  mechanisms  lead to the same result because
the speed of light is constant.  In the  ${M^4}_2$-space  distinctions
between A and B-versions are more essential.  Indeed,  in the case of B
the scattered $\gamma'$-quantum velocity will  be  determined  by  the
expression
\end{sloppypar}
\begin{equation}\label{eq.B}
\displaystyle
c'=c_0\sqrt{1+\frac{{v'}^2}{{c_0}^2}}=
c_0\sqrt{1+\Big(\frac{\hbar\omega}{{\cal E}_0}\Big)^2
\frac{{{\cal E}_0}^2sin^2\theta+({\cal E}_0+\hbar\omega)^2
(1-cos\theta)^2}{[{\cal E}_0+\hbar\omega(1-cos\theta)]^2}}
\end{equation}
instead of  the  formula  (\ref{eq.A}),  because  the Compton electron
velocity (\ref{eq.13}) differs  from  the  virtual  electron  velocity
(\ref{eq.13a}). As  a  result the scattered $\gamma'$-quantum velocity
comes to depend on the  angle  of  its  scattering.  In  the  case  of
forward-scattering with   $\alpha=0,  \  \theta=0$  this  velocity  is
minimal and equal $c'=c_0$.  The $\gamma'$-quantum energy  is  maximal
and coincides,   according   to   (\ref{eq.3}),  with   the   incident
$\gamma$-quantum energy  $\hbar\omega'=\hbar\omega$.  With  the  small
scattering angles              when             $sin\theta\sim\theta$,
$cos\theta\sim(1-\theta^2/2)$ the scattered $\gamma'$-quantum velocity
is        $c'\sim       c_0\{1+(\hbar\omega/{\cal       E}_0)^2[{{\cal
E}_0}^2\theta^2+(\hbar\omega)^2\theta^4/4]/[{\cal
E}_0+\hbar\omega\theta^2/2]^2\}^{1/2}$, if $\hbar\omega\gg{\cal E}_0$.
With $\theta\ll{\cal   E}_0/\hbar\omega$   it   is    equal    $c'\sim
c_0[1+(\hbar\omega\theta/{\cal E}_0)^2/2]\sim  c_0$.  In  the  case of
back-scattering with ($\theta\sim\pi$) the scattered $\gamma'$-quantum
energy is  minimal $\hbar\omega' \sim\hbar\omega/(1+2\hbar\omega/{\cal
E}_0)\sim{\cal E}_0/2$,  but   its   velocity   is   maximal   $c'\sim
c_0(\hbar\omega/{\cal E}_0)$, if $\hbar\omega\gg{\cal E}_0$.

\begin{sloppypar}\noindent
\section{Comparison  of  the  kinematics of Compton - effect within the
${M^4}_1$ and ${M^4}_2$-spaces}
\end{sloppypar}
In sum,   we   can  note  the  following  features  of  Compton-effect
kinematics     within     the     Minkowski     spaces      $({M^4}_1,
{M^4}_2)\subset F^5$.
\begin{itemize}
\item The  expression  for  the  scattered $\gamma'$-quantum frequency
$\omega'$ in the ${M^4}_2$-space coincides with the similar expression
for the  scattered  $\gamma'$-quantum frequency in the ${M^4}_1$-space
(as in SR).
\item The  expressions  for the scattered electron momentum and energy
and  for  the  scattered  $\gamma'$-quantum  momentum  and  energy  in
${M^4}_2$ coincide with the similar expressions within ${M^4}_1$.
\item The distinctions arise in  calculating  the  velocities  of  the
scattered $\gamma'$-quantum  and scattered electron.  Within ${M^4}_1$
the velocity of scattered quantum is always equal  $c_0=3\cdot10^{10}$
cm/s. The Compton electron velocity does not exceed $c_0$.
\item In ${M^4}_2$ in scattering the incident $\gamma$-quantum by  the
immobile  electron  in  the  point of its localization,  the scattered
$\gamma'$-quantum velocity does not depend on the scattering angle and
is equal $c_0=3\cdot10^{10}$ cm/s (as in SR).
\item \begin{sloppypar}In ${M^4}_2$ in emitting the scattered  quantum
by  the virtual electron (version A),  the scattered $\gamma'$-quantum
velocity     is      equal      $c'=c_0\sqrt{1+\hbar^2\omega^2/({{\cal
E}_0}^2+2\hbar\omega{\cal E}_0)}$ and exceeds $c_0$.\end{sloppypar}
\item In ${M^4}_2$ in the case of  spontaneous  transmutation  of  the
virtual electron into the free electron with the following emission of
the scattered   $\gamma'$-quantum   (version    B)    the    scattered
$\gamma'$-quantum  velocity depends on the angle of its scattering and
is equal $c'\sim c_0$ if the scattering occurs forward in the range of
angles $\theta\ll{\cal E}_0/\hbar\omega$, and $c'\sim c_0(\hbar\omega/
{\cal E}_0)$ for the back-scattering.
\item In  ${M^4}_2$  the  Compton  electron  velocity  $v'$  trends to
$\infty$ with $\hbar\omega\to\infty$.  Faster-than-light motion of the
forward-scattered  electron begins from the energy of incident quantum
${\cal E}_0/\sqrt2\sim 360$ keV.
\item In  both  the  Minkowski  spaces the motion of forward-scattered
electron ($\alpha=0$)  corresponds  to   the   motion   of   scattered
$\gamma'$-quantum in  the  backward  direction ($\theta=\pi$) with the
energy $\hbar\omega'=\hbar\omega/(1+2\hbar\omega/{\cal E}_0)\sim {\cal
E}_0/2\sim 250$ keV if $\hbar\omega\gg{\cal E}_0$ (as in SR).
\end{itemize}

\section{Turning to equations of quantum theory}
Let us make  clear  how  the  basic  equations  of   quantum   theory
(Schr\"odinger, Klein-Gordon-Fock and Dirac equations) may be  written
in the  ${M^4}_2$-space.  For  this  purpose we shall use the standard
approach and pass on to the operator form for energy and  momentum  in
the line 6 of Table 1 accordingly to the rule:
\begin{equation}\label{oper}
{\cal E}\to i\hbar(c_0/c)\partial/\partial t, \ \
{\bf p}\to -i\hbar\nabla.
\end{equation}
Here the  operator for energy takes the well-known form \cite{Shir72},
if $c=c_0$.  The operator for momentum is standard \cite{Shir72}. Then
for  the  free  motion  of  a  quantum  particle we have the following
equations.

\begin{sloppypar}{\it The  Schr\"odinger equation}.
\noindent
Taking into consideration that in non-relativistic approximation  with
$v\ll c_0$  the  velocity  of light $c\sim c_0$ and the expression for
kinetic energy $T=p^2/2m$ is the same for ${M^4}_2$ and ${M^4}_1$,  we
find that the Schr\"odinger equation \cite{Shir72} will be the same in
both the spaces:
\begin{equation}\label{schre}
\Big(i\hbar\partial_t+\frac{\hbar^2}{2m}\triangle\Big)\psi(t,{\bf x})=0,
\end{equation}
where $m$  is  the  mass of a particle,  $\psi(t,{\bf x})$ is the wave
function. As a result the non-relativistic quantum theory (the quantum
mechanics)  and the classical mechanics are the same for ${M^4}_2$ and
${M^4}_1$.  The difference will appear in the  relativistic  range  of
motion.  In  ${M^4}_2$  this  range  begins  with  the  energy  ${\cal
E}\geq\sqrt2{\cal E}_{0e}\sim  723$ keV for electron and ${\cal E}\geq
\sqrt2{\cal E}_{0p}\sim 1.33$ GeV for proton. (The velocities of these
particles will be equal or above $c_0=3\cdot 10^{10}$ cm/s).
\end{sloppypar}
{\it The Klein-Gordon-Fock} equation.  With the help of the dispersion
relation (\ref{dis}) we obtain
\begin{equation}\label{klein}
\displaystyle
\Big(\frac{1}{c^2}\partial_{tt}-\triangle+\frac{m^2{c_0}^2}
{\hbar^2}\Big)\Phi(ct,{\bf x})=0.
\end{equation}

{\it The Dirac equation}.
\begin{equation}\label{dirac}
\Big(i\gamma^0\frac{1}{c}\partial_t+i(\gamma^1\partial_x+
\gamma^2\partial_y+\gamma^3\partial_z)-\frac{mc_0}{\hbar}\Big)
\Psi(ct,{\bf x})=0.
\end{equation}
Here $\gamma^0$, $\gamma^1$,  $\gamma^2$,  $\gamma^3$  are  the  Dirac
matrices, $\Phi(ct,{\bf x})$  and  $\Psi(ct,{\bf  x})$  are  the  wave
functions. As  distinct  from  ${M^4}_1$,  in the ${M^4}_2$-space with
$c\to\infty$ the   equations    (\ref{klein}),    (\ref{dirac})    are
characterized by the appearance of solutions not depending on the time
because the components  with  the  derivative  with  respect  to  time
vanish. If $c=c_0$, the equations (\ref{klein}), (\ref{dirac}) go into
SR equations \cite{Akh69, Fus83}.

\section{Discussion and conclusion}
It has been considered the version of a mathematical  theory  that  is
similar  to  SR  but differs from it in view of its being based on the
metric of more general form (\ref{int}).  Here the velocity  of  light
run through     the     continuous    spectrum    of    values    from
$c_0=3\cdot10^{10}$ cm/s to $\infty$.  It is  believed  from  formally
mathematical   standpoint  that  the  space  with  such  a  metric  is
5-dimensional. It  contains  two  Minkowski  space:  the  first  space
${M^4}_1$ on  the  hyper-plane  $c_0$  with the local time $x^0=c_0t$,
where SR  is  realized,  and  the  second  space  ${M^4}_2$  with  the
non-local   time   $x^0=\int_0^t   cd\tau$,   where  realized  is  the
theoretical version  from  the  present  work  and  the   publications
\cite{Kot03} -  \cite{Kot04a}.  Some  like  ideas are contained in the
well-known monograph  of  Pauli  \cite{Pau47}.  On  the  page  29   in
discussing  the  Michelson  experiment,  Pauli notes that according to
Abraham the velocity of light in frame $K'$ moving together  with  the
interferometer is equal
\begin{equation}
c'=c\sqrt{1-\beta^2}.
\end{equation}
This differs  from  the  velocity  of  light $c$ in laboratory frame
$K$\footnote{Abraham, 1908, $\beta=V/c$ \ \cite{Pau47}.}. According to
Abraham  the  time  dilatation  is  absent.  The  Abraham's  viewpoint
conforms to the result of Michelson  experiment  but  contradicts  the
relativity  principle  because  it  leaves  room  for  absolute motion
\cite{Pau47}.
\begin{sloppypar}
It is interesting to note that if we find $c$ from Abraham's formula
and postulate  $c'={c_0}'=3\cdot10^{10}$  sm/s,  we  just  obtain  the
relations  (\ref{time})  and  (\ref{c}) of the present work that is in
agreement with the principle of relativity.  Thus, the Abraham's point
of  view  turned  out  to  be  associated  in an indirect way with the
Finsler space (\ref{int}) and with the presence of the  two  Minkowski
spaces ${M^4}_1$ and ${M^4}_2$ in it.  This is the simplest example of
turning to spaces of such a type.  However this simplicity makes  deep
sense  as  it  is conditioned by fundamental properties of 3-space and
time such as isotropy and homogeneity.  The more complicated  examples
of non-homogeneous        space-time       with       the       metric
$ds^2=c^{-2N}\Big\{[cdt+(1-N)tdc]^2-\sum_j(dx^j-Nx^jdc/c)^2]\Big\}$,
where N  is  the number,  $|c|<\infty$,  $j=1,2,3$,  are considered in
\cite{Kot81, Kot79,  Kot79a}.  This metric enables  one  to  introduce
three Minkowski spaces: on the hyper-plane $c=c_0$=const with the time
$x^0={c_0}^{1-N}t$, on                   the                   vectors
$(x^0=c^{1-N}t,x^j=(c^{-N}x,c^{-N}y,c^{-N}z))$ with     the    time
$x^0=c^{1-N}t$, and   on   the    hyper-plane    $t=t_0$=const    with
$x^0=c^{1-N}t_0$. In  the  last  case  the  role  of  time as a scalar
parameter will play the velocity of light $c$. Motions in this space
will happen beyond the conventional conception of time.  At present it
is not clear what the possibility  of  existing  additional  Minkowski
spaces  means,  as well as whether this possibility has to do with the
physical reality. It is a subject for further investigations.
\end{sloppypar}

In sum,  we  have  shown that in the ${M^4}_2$-space it is possible to
construct  the  theory,  which  admits  faster-than-light  motions  of
electromagnetic  fields and particles with real masses.  As a subgroup
of symmetry,  it  contains  the  Poincar\'e  group.   Unlike   motions
described by SR in ${M^4}_1$, in the ${M^4}_2$-space it is possible to
introduce the time  similar  to  the  universal  Newton  time  on  the
trajectory  of  a  particle.  The particle mass does not depend on the
velocity of its motion and is the fundamental constant as in classical
mechanics. According   to   the   Compton-effect   kinematics  in  the
${M^4}_2$-space the  scattered  electron   will   move   faster   than
$c_0=3\cdot 10^{10}$  cm/s,  if  the  incident $\gamma$-quantum energy
exceeds $360$ keV.  For example,  in  the  case  of  the  annihilation
quantum with the energy  $511$  keV  ($Na^{22}$) and  the  propagation
velocity  $c=c_0$  the  forward-scattered electron will be moving with
the velocity $0.8c_0$ in  the  ${M^4}_1$-space  and  $1.3c_0$  in  the
${M^4}_2$-space.  This  distinction  (if  it  exists  really)  may  be
experimentally detected by means of measuring the flight-time  of  the
Compton  electrons and annihilation $\gamma$-quantum on the base $100$
cm long.
\bigskip

{\bf Acknowledgements}
\smallskip

\noindent The   author   is   deeply   grateful  to  Academician  V.G.
Kadyshevsky  of  RAS  and  Professor  A.E.  Chubykalo   of   Zacatecas
University  for helpful discussion,  valuable and critical remarks and
to Researcher of Kurchatov Institute L.N.  Nefedova for help with  the
work.

\end{document}